\documentstyle[multicol,prl,aps,epsf]{revtex}
\begin{document}
\draft
\title{Anisotropic Coarsening: Grain Shapes and Nonuniversal Persistence}
\author{Andrew D. Rutenberg}
\address{Centre for the Physics of Materials,
McGill University, 3600 rue University, Montr\'{e}al QC, Canada H3A 2T8}
\author{Benjamin P. Vollmayr-Lee}
\address{Department of Physics, Bucknell University, Lewisburg, PA 17837}
\date{March 30, 1999}
\maketitle
\begin{abstract}
We solve a coarsening system with small but arbitrary anisotropic surface 
tension and interface mobility. 
The resulting size-dependent growth shapes are significantly different from 
equilibrium microcrystallites, and have a distribution of grain sizes 
different from isotropic theories.  As an application of our results, we
show that the persistence decay exponent depends on anisotropy 
and hence is nonuniversal. 
\end{abstract}

\begin{multicols}{2}

The geometrical Wulff construction \cite{Wulff01} gives an explicit
relation between the anisotropic surface tension and the 
resulting equilibrium crystal shape.  This marks an early and
dramatic success in quantitatively connecting morphology to the 
interfacial properties of a material.  However, distinct Wulff 
microcrystallites must be in `splendid isolation' --- with negligible 
exchange between them in comparison to the internal dynamics required 
to equilibrate \cite{Wortis88}.  
In contrast, dilute phase separating alloys and coarsening polycrystallites
exhibit growing microcrystalline droplets or grains with 
non-negligible interactions.  
While it has been shown for these and other coarsening systems
that anisotropy influences the
morphology \cite{Rutenberg96,Holm91}, such effects 
have not been quantitatively understood for even the simplest models of 
curvature-driven growth.  

The understanding of interacting {\em isotropic} phases 
(see, e.g., \cite{Bray94})
was significantly advanced by the models of
Lifshitz and Slyozov \cite{Lifshitz59} and Wagner \cite{Wagner61} for
diffusive and curvature-driven coarsening, respectively.
These mean-field theories correctly capture a remarkable amount of 
coarsening phenomenology, and are exact in the dilute limit. 
With this inspiration, we generalize Wagner's model ---
an interacting ensemble of coarsening droplets, evolving to continually lower 
their surface energy without changing their total volume ---
to include arbitrary anisotropy in the surface tension and
the interface mobility. 
We solve the model perturbatively in anisotropy strength, and relate the
interfacial properties to the resulting non-trivial grain 
shapes.  These ``growth shapes'' are contrasted with those of equilibrium
(Wulff-constructed) grains to highlight the connection between dynamics 
and microcrystallite morphology.  We then compare our results
on the ensemble of grains
to Wagner's isotropic solution to demonstrate anisotropy effects on
coarsening correlations, including the effect on persistence exponents.

Our model is applicable to single-phase polycrystallite coarsening,
where distinct grains are distinguished only by their crystallographic 
orientation  (see \cite{Adams98,Glazier92,Mullins86}). Most 
theoretical studies of polycrystallites focus on
their cellular structure, specifically on the static and 
dynamical description of the vertices where three or more grain boundaries 
meet.  However, 
vertex-based models have significant shortcomings when anisotropy
is included, since it modifies both the distribution of the number
of vertices per grain \cite{Holm91} and the otherwise fixed
angles formed where three grains adjoin \cite{Adams98}. Furthermore, 
von Neumann's law, a direct relationship in two dimensions (2D)
between the number of vertices per grain and its area growth rate 
\cite{Glazier92}, 
no longer applies. With anisotropy, the evolution of a grain's area 
requires the complete specification of grain shape --- including the 
orientations and, in general, the  non-uniform curvatures of the interfaces. 

We present a complementary vertex-free approach to examine grain shape 
via  an anisotropic dynamical mean-field theory.
The neighboring grains outside the grain of interest are
treated as providing an isotropic mean-field. 
We retain the crystallinity of the grain
through an anisotropic surface tension and interface mobility, which
results in the anisotropic Wagner theory.
(A similar connection can be made between isotropic Wagner theory and 
soap froths \cite{Sire95}.)   Ultimately, a synthesis of the present 
work with vertex-based models is desirable \cite{vertices}.

We find a dynamical scaling solution typical of coarsening 
systems \cite{Bray94}, including clean polycrystallites.  
The characteristic length scale grows as a
power law, $L \sim t^{1/2}$, as expected for curvature driven growth
\cite{Mullins86}. In the scaling regime the initial conditions are
``forgotten,'' and the morphology, when scaled by the growing length, 
$L(t)$,  is invariant.  Grain shapes of particular scaled size 
are also time-independent. 
These growth shapes are generally quite different from 
equilibrium Wulff shapes --- even when the mobility is isotropic! 
The isotropic grain size distribution is also modified by anisotropy, as
discussed later. 

With our results, we can answer the question 
of universality in persistence decay exponents. The persistence is
the fraction of the system that has not been crossed by a domain wall up to
time $t$ \cite{persistence,Lee97,Majumdar98}.
The decay of persistence to zero, $P \sim t^{-\theta}$, even from a starting
time deep within the scaling regime, implies 
that every point in the system will eventually ``realize'' that equilibrium 
has not yet been reached.  Persistence decay is a {\em local} signature of 
the non-equilibrium dynamics of the system.  The degree of universality of 
this dynamical exponent has remained an open issue since no precise results 
have been obtained for models with non-trivial temperature dependence 
\cite{trivtemp}.  (Simulations have not yet found any 
temperature dependence within their 
accuracy \cite{Cueille97}.) Since anisotropy varies with temperature, 
our model provides such 
a non-trivial temperature dependence in a coarsening system
that we can then analytically relate to the resulting
structure and to the persistence exponent, $\theta$ \cite{Lee97}. 
We find that $\theta$ depends on both the anisotropies of 
the surface tension and of the interface mobility, so the persistence 
exponent is nonuniversal in anisotropic systems \cite{universality}.

We restrict ourselves to 2D, where the
surface tension $\sigma(\psi)$ and the interface mobility $M(\psi)$ may be
defined in terms of the angle $\psi$ between the interface normal and an
arbitrary crystallographic axis.
The anisotropic Allen-Cahn equation \cite{Siegert90,McFadden93} is then
derived from the linear response of the interface to the local 
drive given by the Gibbs-Thompson condition,  
$[\sigma(\psi)+\sigma''(\psi)] \kappa$, were  $\sigma$ is the surface
tension, and $\kappa$ the local interface
curvature.  The stiffness, $\sigma +\sigma''$, reflects  the
local change of extent {\it and\/} orientation of the interface due to a 
deformation.  By allowing the interface mobility to depend
on orientation \cite{mobility}, and by including an applied field $\lambda$ 
coupled to one of the phases, we obtain the normal interface velocity
\begin{equation}
\label{EQN:vn}
   v_n = - M(\psi) \left[
           \{ \sigma(\psi)+\sigma''(\psi) \} \kappa - \lambda \right].
\end{equation}

\narrowtext
\begin{figure}
\centerline{\epsfxsize = 2.4truein
\epsfbox{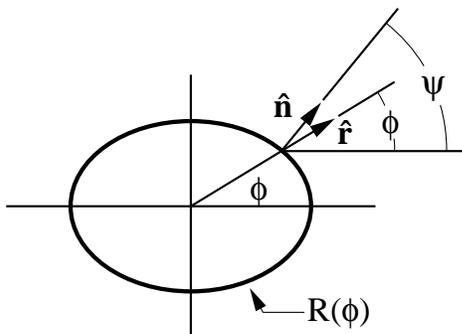}}
\caption{An anisotropic drop illustrating $\phi$, the
polar coordinate, and $\psi(\phi)$, the angle of the interface normal
at the point $\bigl(R(\phi),\phi\bigr)$.  
\label{FIG:psiphi}}
\end{figure}

We now consider an ensemble of polycrystallite grains. Our mean-field
approximation entails keeping only the crystalline anisotropy 
of each grain (ignoring its neighbors), neglecting vertices, 
and determining a self-consistent mean-field $\lambda$ to represent the 
effects of neighboring grains that may be growing or shrinking. The 
conservation of the total area of all of the grains uniquely 
determines $\lambda(t)$, resulting in precisely the anisotropic 
Wagner theory.

To proceed, we Fourier expand the anisotropic surface
tension and mobility, 
\begin{eqnarray}
\label{EQN:sigma}
     \sigma(\psi) &=& \sigma_0 \Bigl[1 + \delta \sum_{k=1}^\infty
             \{\sigma_k \cos(k\psi) + \tilde\sigma_k \sin(k\psi)\} \Bigr], \\
\label{EQN:M}
     M(\psi) &=& M_0 \Bigl[1 + \delta \sum_{k=1}^\infty
              \{m_k \cos(k\psi) + \tilde m_k \sin(k\psi) \}\Bigr],
\end{eqnarray}
where $\delta$ is introduced to organize a perturbative calculation.
We parameterize each grain by a polar
radius $R(\phi)$, as depicted in  Fig.~\ref{FIG:psiphi}, from which
the interface orientation follows: $ \psi(\phi) = \phi - \arctan(R'/R)$ where
$R' \equiv dR/d\phi$.   Considering only smooth grain profiles,
we relate normal and radial growth velocities, 
$v_r(\phi) = v_n \sqrt{1+ (R'/R)^2}$, and calculate the curvature 
$\kappa(\phi) = [R^2 + 2 R'^2 - R R'']/(R^2+R'^2)^{3/2}$.
We then expand $R$, 
\begin{equation}\label{EQN:Rexp}
    R(\phi) = R_0 \Bigl[1+ \sum_{k=0}^\infty 
                \{ \rho_k \cos(k\phi) +  \tilde \rho_k \sin(k\phi)\} \Bigr],
\end{equation}
with coefficients 
\begin{equation}\label{EQN:rhoexp}
    \rho_k(x) = a_k(x) \delta + b_k(x) \delta^2 + \dots
\end{equation}
and similarly for $\tilde{\rho}_k$ \cite{origin}. 
Grain sizes are labeled with a
reduced length $x \equiv R_0/L$, where $L\equiv (M_0 \sigma_0 t/2)^{1/2}$.
For $\delta=0$ we recover Wagner's isotropic theory, with the
familiar distribution of grain sizes (see \cite{Wagner61,Sire95,Lee97}):
\begin{equation}
\label{EQN:isotropic}
         f(x) = \epsilon F_2 x \exp[-4/(2-x)]/(2-x)^4.
\end{equation}
(The $\epsilon$ prefactor is the area fraction of a randomly selected
subset of grains --- used later to calculate persistence.)
For convenience, we {\em define} $R_0$ by the requirement that $x$ maintains
this {\em isotropic} grain-size distribution
up to an anisotropy-dependent normalization, $F_2 = F_2^{(0)}+
\delta^2 F_2^{(2)}+\ldots$. 
This requirement leads to  non-zero
$\rho_0$ terms in the expansion (\ref{EQN:Rexp}) but 
preserves the range of scaled sizes, $x \in [0,2]$.
(Note that $\tilde \rho_0=0$.) Physical length scales, 
such as the grain perimeter, can be consistently 
derived from our results, as discussed below.

\begin{figure}
\centerline{\epsfxsize = 3.2truein
\epsfbox{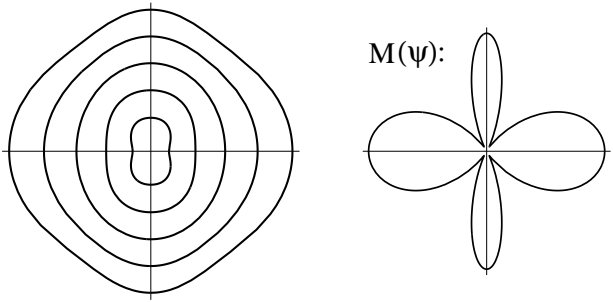}}
\caption{First order grain shapes for various sizes (not to scale) with
an isotropic surface tension $\sigma_0$ but with a particular 
anisotropic mobility $\delta m_2=0.4$ and $\delta m_4=0.9$ (all other
$m_k =\tilde m_k=0$).  The
scaled grain sizes are, from the innermost, $x=0.01, 0.5, 1, 1.5$, and $1.99$.
For no value of $x$ is there a circular grain, the equilibrium Wulff shape.  
Angles for which $M(\psi)$ is larger correspond 
roughly to larger radius in growing (larger) grains, and smaller radius
in shrinking (smaller) grains.
\label{FIG:s0}}
\end{figure}

The resulting interface equations for the ensemble of grains may be solved
order by order in $\delta$ \cite{long}. 
The zeroth order results reproduce the isotropic theory; the first order 
equations are new, and serve to determine a size-dependent grain shape
through $a_k(x)$:
\begin{eqnarray}
\label{EQN:ak}
                x(2-x)^2 a_k'(x) - 4 (k^2+x-2) a_k = \nonumber \\ 
                                4(1-k^2)\sigma_k+ 4(1-x) m_k,
\end{eqnarray}
for $k>1$, with an identical equation for $\tilde a_k$ in terms of 
$\tilde\sigma_k$ and $\tilde m_k$.  For $k>1$ the solution is 
\begin{equation}
\label{EQN:akintegral}
                a_k(x) = m_k + (\sigma_k-m_k) (1-1/k^2) [1+\Omega(k,v)]
\end{equation}
where $\Omega(k,v) \equiv  2 \Gamma(2-k^2,v) v^{k^2-2} e^v$, and 
$v \equiv k^2 x/(2-x)$.  We also have $a_0(x)=a_1(x)=0$, 
the latter by our choice of coordinate origin \cite{origin}.  Clearly
the grain shapes depend on grain size, through $\Omega$.  Even when 
the surface tension is isotropic ($\sigma_k=0$ for all $k>0$), we can
obtain anisotropic grain shapes through the interface mobility.
This is illustrated in Fig.~\ref{FIG:s0} for a particular choice of
$M(\psi)$. 

At all orders of $\delta$ the equations for the grain shape are similar
to (\ref{EQN:ak}), although the right-hand side will include products 
of lower-order solutions.  While these equations are 
progressively more difficult to solve, we can 
iteratively demonstrate that the solutions are finite at every
order of $\delta$ \cite{long}.  

In the special case where $m_k=\sigma_k$ for all $k$ 
grains of all sizes have the equilibrium Wulff shape.
This result holds {\em to all orders} in $\delta$, and 
is due to a remarkable symmetry held by the interface equation 
(\ref{EQN:vn}).  The equilibrium grain shape is given by 
\begin{equation}
\label{EQN:wulff}
        R_{\rm eq}(\phi) = \frac{R_0}{\sigma_0} \min_{\phi'}
                    \left| \frac{\sigma(\phi')}{\cos(\phi'-\phi)} \right|.
\end{equation}
For this Wulff shape, 
a variational calculation shows that $[\sigma(\psi)+\sigma''(\psi)]\kappa$
is independent of angle \cite{long}, from which we obtain 
$v_r \propto M(\psi)\sqrt{1+(R'/R)^2}$ for all angles.  If and  only
if  the {\it dynamical\/} 
mobility anisotropy equals that of the {\it static\/} surface tension ---
that is, $M(\psi) \propto \sigma(\psi)$ ---
then we recover $v_r \propto R_{\rm eq}(\phi)$,  
the condition for Wulff grains to keep their shape while evolving.
This symmetry, evident in (\ref{EQN:akintegral}), 
leads to size-independent drop shapes and also shows up in 
the drop size  distribution and persistence results, as discussed below.  

However, the dynamic mobility and the static
surface tension will {\em not} be proportional except by special construction.
Regardless, in physical systems $M$ and
$\sigma$ have different temperature dependences so that equality could
not be maintained as temperature varies.  In the general case, we will
have size-dependent drop shapes given, to first order, by 
(\ref{EQN:akintegral}). In comparison, the Wulff construction gives 
$a_k^{\rm eq}=\sigma_k$ as the leading contribution to the equilibrium grain 
shape.   Even with an isotropic mobility, $m_k=0$, growth
shapes differ from equilibrium and depend on grain size. 

The isotropic grain size distribution (\ref{EQN:isotropic}) applies only
to our index $R_0$ \cite{origin}.  Physically relevant 
lengths, such as extracted from the grain perimeter or the area, will generally
have different distributions.
For example, the area $A={1\over 2}\int_0^{2\pi}d\phi\,R(\phi)^2$ 
can be used to define $R_A=\sqrt{A/\pi}$ where
\begin{equation}
        R_A = R_0 \Bigl[ 1 + \delta^2 \Bigl( b_0 + {\textstyle {1\over 4}}
                \sum_k \{a_k^2 +\tilde a_k^2 \} \Bigr) + O(\delta^3)
                        \Bigr]. 
\end{equation}
A scaled size $z=R_A/L$ may then be introduced, which will be related
to $x$ by $z=x+\delta^2 h(x) + O(\delta^3)$. The ``area radius'' distribution
$g(z)$ is then determined by $g(z)dz=f(x)dx$ so that 
\begin{equation}
        g(z) = f(z) - \delta^2 [f(z)h'(z) + f'(z)h(z)] + O(\delta^3)
\end{equation}
where $f(x)$ is the isotropic distribution \cite{long}.
The grain perimeter distribution follows similarly, though with a different
function $h(x)$.  [In the special symmetric case, where $m_k=\sigma_k$ for
all $k \geq 1$, all physical lengths have the same distribution.  Since the 
grain shapes are size independent, $h(x)=h_0$, and
$g(z)$ and $f(x)$ differ only by an overall normalization.]

We may also calculate the slow decay of persistence due to the evolution of a
small area fraction $\epsilon$ of randomly chosen grains, following
\cite{Lee97,long,Tam97}. The persistence $P_>$ of the region outside
the chosen grains decays due to growing
grains via $\partial_t P_> = -v_> P_>$. The rate of encroachment of
growing grains, $v_>$, can be calculated from the grain shapes and 
(\ref{EQN:vn}). The power-law decay of persistence follows
directly from the result $v_>\propto 1/t$, with persistence exponent 
$\theta = t v_>$.  Anisotropy appears
at $O(\delta^2)$.  The calculation is lengthy and details are reported 
elsewhere \cite{long}; however, the result simplifies to
\begin{equation}
\label{EQN:theta}
        \theta = \theta_0 + \delta^2 \sum_{k=1}^\infty \theta_k^{(2)}
        [(m_k-\sigma_k)^2 + (\tilde m_k-\tilde\sigma_k)^2] 
        + O(\delta^3),
\end{equation}
where $\theta_0 \simeq 0.48797\epsilon$ is the 2D
persistence exponent for the isotropic case \cite{Lee97}.
We find that $\theta$ equals the isotropic value
$\theta_0$ only when $M(\psi) \propto \sigma(\psi)$ 
(this holds to all orders due the 
symmetry mentioned earlier) and differs from $\theta_0$ for any other
anisotropic conditions.
The order $\epsilon$ coefficients $\theta_k^{(2)}$ may be determined by 
numerical integration, and are well 
approximated by a large $k$ expansion \cite{largek}.
The persistence exponent depends
continuously on both the mobility and surface tension, and consequently on 
the temperature.  The 2D Ising model provides an
explicit example: the anisotropic surface 
tension is known analytically for $0 \leq T \leq T_c$ \cite{Abraham77}
and the anisotropic mobility is known close to $T=0$ for
Glauber dynamics \cite{Spohn93}.  At $T=0$, and using only
the leading contribution (\ref{EQN:theta}), we find
$\theta\simeq 1.0344 \; \theta_0$.  The effect on the exponent is small but 
non-zero, and may be detectable numerically with more accurate studies.

In conclusion, we have constructed a mean-field model for 2D
polycrystallite coarsening with anisotropic surface
tension and mobility. We find an exact scaling solution with 
size-dependent grain shapes that are
generally unrelated to the equilibrium Wulff shape. We use our
solution to calculate the exponent describing persistence decay, and find
that it is {\em continuously dependent on anisotropy} and hence
nonuniversal with respect to temperature \cite{universality}. 
We expect similar results to hold in three-dimensional systems.

We hope that this study stimulates further research of the connections
between nonequilibrium structure and anisotropic mobility and surface tension.
Our next step will be to develop the anisotropic generalization of 
Lifshitz-Slyozov diffusive coarsening in bulk systems \cite{Lifshitz59}.
We also feel the influence of anisotropy on nonequilibrium 
exponents needs further study.  For example, we suspect that persistence 
exponents and their various generalizations (see, e.g., \cite{Majumdar98}) 
will prove to be nonuniversal whenever correlation functions are anisotropy
dependent.
Finally, we stress that persistence decay still provides an important
description of nonequilibrium dynamics, and remains universal in intrinsically
isotropic systems such as binary fluids, polymer blends, and soap froths. 
Persistence is particularly useful in discriminating between different 
dynamical models and in probing the dynamics of soap froths \cite{Lee97,Tam97}.

We would like to acknowledge stimulating discussions with 
S. N. Majumdar, B. Meerson, C. Carter and J. Warren.
A. D. R. thanks the NSERC, and {\it le Fonds pour la Formation de Chercheurs 
et l'Aide \`a la Recherche du Qu\'ebec} for financial support;
B. P. V.-L. was supported by an NRC Research Associateship for part of 
this work.

\end{multicols}

\begin{references}

\bibitem{Wulff01} G. Wulff, Z. Kristallogr.\ Mineral.\ {\bf 34}, 449 (1901);
more recently, see also \protect\cite{Wortis88}.

\bibitem{Wortis88} M. Wortis  in {\em Chemistry and Physics of Solid
Surfaces VII}, edited by R. Vanselow and R. F. Howe (Springer-Verlag,
Berlin, 1988), p.\ 367.

\bibitem{Rutenberg96} A. D. Rutenberg, Phys.\ Rev.\ E {\bf 54}, 2181 (1996).

\bibitem{Holm91} E. A. Holm, J. A. Glazier, D. J. Srolovitz, and G. S.
Grest, Phys. Rev. A {\bf 43}, 2662 (1991).

\bibitem{Bray94} A. J. Bray, Adv.\ Phys.\ {\bf 43}, 357 (1994). 

\bibitem{Lifshitz59} 
I. M. Lifshitz and V. V. Slyozov, Zh.\ Exp.\ Teor.\ Fiz.\ {\bf 35}, 479 (1958);
    [Sov.\ Phys JETP.\ {\bf 8}, 331 (1959)];  ibid., J. Phys.\ Chem.\ Solids
        {\bf 19}, 35 (1961).

\bibitem{Wagner61} C. Wagner, Z. Elektrochem.\ {\bf 65}, 581 (1961).

\bibitem{Adams98} B. L. Adams, D. Kinderlehrer, W. W. Mullins, A. D.
Rollett, and S. Ta'asan, Scripta Metall. {\bf 38}, 531 (1998).

\bibitem{Glazier92} J. A. Glazier and D. Weaire, J. Phys.\
  Cond.\ Matt.\ {\bf 4}, 1867 (1992), and references therein.

\bibitem{Mullins86} W. W. Mullins, J. Appl. Phys. {\bf 59}, 1341 (1986), see
also, e.g., E. L. Holmes and W. C. Winegard, Acta. Metall. {\bf 7}, 411
(1959). 

\bibitem{Sire95} C. Sire and S. N. Majumdar, Phys.\ Rev.\ Lett.\ {\bf 74},
        4321 (1995); {\em ibid.}, Phys.\ Rev.\ E {\bf 52}, 244 (1995). 

\bibitem{vertices} Correlations between vertex locations and the
crystalline axes could affect average grain shape. 
More significantly, the {\em average} grain shape, when the distributed force
due to vertices is included, does not necessarily evolve by curvature-driven
growth.

\bibitem{persistence}
   Persistence was introduced in A. J. Bray, B. Derrida, and C. Godr\`eche, 
   Europhys.\ Lett.\ {\bf 27}, 175 (1994); M. Marcos-Martin {\it et al.}, 
   Physica A {\bf 214}, 396 (1995).  See \protect\cite{Lee97} and
   \protect\cite{Majumdar98} for recent references.

\bibitem{Lee97} B. P. Lee and A. D. Rutenberg, Phys.\ Rev.\ Lett.\ {\bf 79},
4842 (1997).

\bibitem{Majumdar98} S. N. Majumdar and A. J. Bray, Phys.\ Rev.\ Lett.\ {\bf
81}, 2626 (1998).

\bibitem{trivtemp} For isotropic systems, it is believed that 
the temperature only modifies the global kinetic coefficient or 
time scale, and not the scaled structure \cite{Bray94}. 

\bibitem{Cueille97} S. Cueille and C. Sire, J. Phys.\ A {\bf 30}, L791
(1997); ibid cond-mat/9803014; 
B. Derrida, Phys.\ Rev.\ E {\bf 55}, 3705 (1997).

\bibitem{universality} Such behavior is consistent with a renormalization
group description where the fixed point depends on the coarse-grained
$\sigma(\psi)$, $M(\psi)$, and area fraction $\epsilon$.  However,
the universality classes are impractically small,
 in contrast to the behavior of growth exponents.

\bibitem{Siegert90} M. Siegert, Phys.\ Rev.\ A {\bf 42}, 6268 (1990).

\bibitem{McFadden93} G. B. McFadden, A. A. Wheeler, R. J. Braun, S. R.
Coriell, and R. F. Sekerka, Phys.\ Rev.\ E {\bf 48}, 2016 (1993).

\bibitem{mobility} The {\it interface\/} mobility generically differs from
the order-parameter mobility as it depends not only on the order-parameter 
dynamics, but also on the interface profile.  See \cite{long}.

\bibitem{origin} To uniquely determine the origin, 
we require $\rho_1=\tilde{\rho}_1=0$.

\bibitem{long} B. P. Vollmayr-Lee and A. D. Rutenberg, unpublished. 

\bibitem{Tam97} W. Y. Tam {\em et al.}, Phys.\ Rev.\ Lett.\ {\bf 78}, 1588
        (1997);  W. Y. Tam, A. D. Rutenberg, K. Y. Szeto, and 
                B. P. Vollmayr-Lee, in preparation.

\bibitem{largek} $\theta_k^{(2)}/\epsilon \approx 0.0912624
  -0.3035888/k^2+0.2744548/k^4+0.189615/k^6$
   is accurate to within $1\%$ at $k=2$ and $0.004\%$ for $k \geq 4$.
\bibitem{Abraham77} D. B. Abraham and P. Reed, J Phys.\ A {\bf 10}, L121
(1977); J. E. Avron, H. van Beijeren, L. S. Schulman, and R. K. P.
Zia, J. Phys.\ A {\bf 15}, L81 (1982).

\bibitem{Spohn93} H. Spohn, J. Stat.\ Phys.\ {\bf 71}, 1081 (1993).

\end{references}
\end{document}